\documentclass[12pt,rmp,twocolumn,natbib,amsmath,amssymb,showkeys,pre,floatfix]{revtex4}
\usepackage[dvips]{graphicx}

\begin{document}

\title[]
{Ecological Complex Systems}

\author{\firstname{Astero} \surname{Provata$^{\sharp}$}}
\author{\firstname{Igor M.} \surname{Sokolov$^{\flat}$}}
\author{\firstname{Bernardo} \surname{Spagnolo$^{\star}$}}
\affiliation{$^{\sharp}$ Institute of Physical Chemistry, National
Center for Scientific Research \\ "Demokritos", 15310 Athens, Greece
\footnote{e-mail: aprovata@chem.demokritos.gr}}
\affiliation{$^{\flat}$Institut f¨ur Physik, Humboldt-Universit¨at zu Berlin,\\
 Newtonstr. 15, D-12489 Berlin \footnote{e-mail: igor.sokolov@physik.hu-berlin.de}}
\affiliation{$^{\star}$ Dipartimento di Fisica e Tecnologie Relative
and CNISM-INFM,\\ Group of Interdisciplinary
Physics\footnote{http://gip.dft.unipa.it}, Universit$\grave{a}$ di
Palermo, \\ Viale delle Scienze, I-90128, Palermo,
Italy\footnote{e-mail: spagnolo@unipa.it}}

\begin{abstract}

Main aim of this topical issue is to report recent advances in noisy
nonequilibrium processes useful to describe the dynamics of
ecological systems and to address the mechanisms of spatio-temporal
pattern formation in ecology both from the experimental and
theoretical points of view. This is in order to understand the
dynamical behaviour of ecological complex systems through the
interplay between nonlinearity, noise, random and periodic
environmental interactions. Discovering the microscopic rules and
the local interactions which lead to the emergence of specific
global patterns or global dynamical behaviour and the noise's role
in the nonlinear dynamics is an important, key aspect to understand
and then to model ecological complex systems.
\end{abstract}
\date{\today}
\maketitle

\section{Introduction}
Ecological complex systems are open systems in which the
interactions between the constituent parts are nonlinear and the
interaction with the environment is noisy. These systems are
therefore very sensitive to initial conditions, deterministic
external perturbations and random fluctuations always present in
nature. The study of noisy non-equilibrium processes is fundamental
for modeling the dynamics of ecological systems and for
understanding the mechanisms of spatiotemporal pattern formation in
ecology. Recently, considerable effort has been devoted to gain
understanding of how population fluctuations arise from the
interplay of noise, forcing, and nonlinear dynamics. The
understanding of complexity in the framework of ecological systems
may be enhanced starting from the "so called" simple systems, in
order to catch the phenomena of interest, and adding details that
introduce complexity at many levels. In general, the effects of
small perturbations and noise, which is ubiquitous in real systems,
can be quite difficult to predict and often yield counterintuitive
behavior. Even low-dimensional systems exhibit a huge variety of
noise-driven phenomena, ranging from a less ordered to a more
ordered system
dynamics~\cite{Sci99,Sci01,Fre02,Spa04,Ebe05,Chi05,Spa08,Liu08}.
\vspace{-0.1cm}

In the past years the study of deterministic mathematical models of
ecosystems has clearly revealed a large variety of phenomena,
ranging from deterministic chaos to the presence of spatial
organization. These models, however, do not account for the effects
of noise despite the fact that it is always present in actual
population dynamics. Frequently, noise effects have been assumed to
be only a source of disorder. Now researchers are gaining a better
understanding of population dynamics by bringing noise into
nonlinear models, both for marine and for terrestrial
animals~\cite{Sci99,Sci01,Fre02,Spa04,Ebe05,Liu08}. A strong
interaction between animals and their environment is responsible for
the damsel-fish cycle and for population fluctuations of feral sheep
for example. Because their survival depends on several interacting
factors, the fish populations can react dramatically to what looks
like small amounts of noise. Damsel-fish and feral sheep populations
are only two examples of a growing list of ecological systems in
which nonlinear dynamics seems to amplify the effects of noise
caused by environmental forcing: Wind, waves, and the phase of the
moon interact, in a multiplicative way, to produce drastic swings in
fish population dynamics. The noise through its interaction with the
nonlinearity of the living systems can give rise to new,
counterintuitive phenomena like noise-enhanced transport,
noise-sustained synchronization, noise-induced transitions,
noise-enhanced stability, stochastic resonance, noise delayed
extinction, temporal oscillations and spatial patterns. In addition,
the analysis of experimental data of population dynamics frequently
needs to consider spatial heterogeneity. Characterizing the
resulting spatiotemporal patterns and the spatial organization is,
perhaps, the major challenge for ecological time series analysis and
for dynamics modeling~\cite{Lev04}.

\vspace{-0.09cm} The collection of papers of this volume gives a
snapshot of the present status of this very interesting and rapidly
expanding interdisciplinary research field, after the workshop on
"\emph{Ecological Complex Systems: Stochastic Dynamics and
Patterns}" held in Terrasini (Palermo, Sicily) from 22 to 26 July,
2007. The workshop had its focus on the comprehension of the noise
role and the formation of spatiotemporal patterns in the dynamics of
ecological complex systems. The lectures of the small scale and
intense workshop had a marked interdisciplinary character. The
workshop gathered 40 participants, including 23 speakers, working at
Universities and Research Institutions of Israel, Germany, Greece,
UK, Poland, Spain, Switzerland, Russia, USA and Italy. The workshop
brought together theoreticians and experimentalists, to establish a
common language and exchange new ideas and recent developments in
the field of ecology with the specific focus on the stochastic
dynamics and spatiotemporal pattern formation in population
dynamics. The discussion after each presentation made the workshop
into a lively, interesting scientific event with a real stimulating
and participating atmosphere. Particularly it is worthwhile to
mention the round table on cancer growth dynamics, which was
characterized by a passionate discussion about open problems on this
interdisciplinary research subject, with an insight into cancer
biology. This round table was a sort of synthesis of all previous
sessions: By discussing firstly on intra and inter gene problems,
then by analyzing ecosystems and finally by ending with cancer
growth dynamics. The focus was on how Physics can help Biology and
vice versa how Biology and Medicine can stimulate and provoke
Physics with their complexity.

\par This topical issue contains a selection of peer-reviewed papers of
the presented contributions at the Workshop, which we shortly
review. Specifically the themes addressed in this issue are:
self-organization and spatiotemporal patterns, bioinformatics,
population dynamics, motion of proteins, motor protein and
interdisciplinary physics.

\section{Spatiotemporal patterns}

Natural ecological structures on earth range in spatial scales from
a few millimeters up to kilometers and in temporal scales from a few
milliseconds to centuries. They are results of natural activity and
competition between different species. On the small scale spatial
pattern formation one can mention the bacterial
colonies~\cite{Ben06} and tumor growth~\cite{Kha06}, cell division
and multiplication, followed in intermediate scales by insect nest
and path patterns (ants, wasps, fish, etc..)~\cite{The02}. At large
scales these are rainforest patterns, which cover many kilometers in
space. At different temporal scales we list self-organized motion
patterns in groups of animals~\cite{Cou05,Erd05}, oscillations in
epidemic spreading and in predator-prey systems~\cite{Pek06} and
stochastic resonance phenomena in protein systems. To describe
complex ecosystems, therefore, it is fundamental to understand the
interplay between noise, periodic and random modulations of
environment parameters, the intrinsic nonlinearity of ecosystems and
to comprehend spatiotemporal dynamics~\cite{Ebe05}.
\vspace{-0.1cm}

It is now becoming apparent that fluctuations and noise are
essential ingredients of life processes. An emerging area of
population dynamics where the interplay between nonlinearity and
noise is essential for the outcome of the process is the virus
dynamics. The interactions between viruses or other infectious
agents and immune system cells together with the interaction with
the noisy environment play a central role in any basic understanding
of the evolution of virulence~\cite{San03,Gro06}. Modeling the
dynamics of cancer growth by taking into account the noise is
another emergent area of research in the field of biological
complexity~\cite{Kha05,Fia06}. The relative rate of neoplastic cell
destruction, for example, is a random process due to the action of
cytokines on the immune systems and could be modeled as a random
dichotomous noise in models of tumor growth. The elaboration of new
models for cancer growth dynamics is a challenge for theoretical
physics and at the same time may lead to new experiments in this
subject area. The problem of target search of the specific binding
site on a genome by a transcription factor~\cite{Sok05} and that of
the mean first passage time analysis to describe the dynamics of the
complex free energy landscape of protein folding are two open
problems with high research impact in ecological complex
systems~\cite{Wan04,Li08}.

\par To study ecological dynamics and their spatiotemporal structures
numerical approaches have been used in order to address the specific
mechanisms which are responsible for the emerging patterns (spatial,
temporal or both). Early numerical studies used mean field nonlinear
models, with notable example the Lotka-Volterra model and its many
off-springs, to address the spatial and temporal complexity of
predator-prey systems. Similar birth-and-death mean field models
were introduced for the study of epidemics, growth models etc... In
recent studies, it was realized that mean field models only
partially can represent the processes taking place on system in
finite dimensions (1, 2, 3 or fractal) and that the finite size of
interactions has to be seriously taken into account for a realistic
representation of ecological structures. In this respect, discreet
ecological models were devised which take into account the specific
neighborhood environment, and factors such as noise, external
conditions acting locally on the system, local diffusion rules,
supports of specific geometry (including fractal supports)
etc...~\cite{Aba03}. Ideas borrowed from statistical mechanics,
nonlinear dynamics, critical phenomena, non-equilibrium physics and
chemistry together with detailed simulations have considerably
contributed to the understanding of ecological pattern formation
processes and its complex dynamics.

\subsection{Self-Organization}

Amplifying communications are a ubiquitous characteristic of
group-living animals. The self-organizing patterns associated with
resource exploitation in social insects have been addressed
theoretically in the paper by Nicolis and Dussutour (p. 379).
Emphasis was placed on the transition from individual behaviour
where only local information is available, to the collective scale
where the colony as a whole becomes capable of choosing between the
different options afforded by the multiplicity of the solutions of
the underlying evolution laws. The basic mechanisms underlying the
phenomena to be considered are: (a) competition between different
sources of information, and (b) the occurrence of amplifying
interactions between constituent units as reflected by the presence
of feedback loops. In particular the response of the ecosystem may
switch between the options available in an unexpected way, depending
on the initial conditions and the ranges of parameter values
considered. A mean field approach incorporating the principal
sources of cooperativity was used and complemented by stochastic
simulations through Monte Carlo approach. Although obtained in the
context of social insect biology, the results of this paper are in
many aspects paradigmatic, and they are expected to apply to a
variety of other biological processes~\cite{Cam01} or artificial
systems.

\vspace{-0.09cm} There are two basic types of models designed for
Lotka- Volterra systems simulation. The first one is a group of
phenomenological macroscopic models based on the ODE description
(for example mean-field (MF) models). The second type of models
based on various stochastic cellular automata and similar numerical
methods provide the direct microscopic simulation of the underlying
processes. The Kinetic Monte Carlo (KMC) simulation is one of the
most popular among them~\cite{Zha99}. KMC enables one to consider
all aspects of spatial and temporal dynamics of the system and
therefore can serve as a more adequate tool for the analysis of
ecological models. A Lattice Lotka-Volterra (LVV) model on a square
lattice under external influence called "long range mixing" is
presented in this issue by Shabunin and Efimov (p. 387). It consists
of a random shuffling of the species across the lattice space. Such
a "mixing" increases the uniformity of the units distribution on the
lattice and changes the character of spatio-temporal processes on
it. The LLV system on a square lattice with local random
interconnections can be synchronized by external long-range mixing
of sufficiently small intensity. This phenomenon has a threshold
character: there is a critical value of the mixing force after which
the behavior of the system is dramatically changed. On the
macro-level it is observed as birth of periodic oscillations after a
super-critical Hopf bifurcation. The amplitude of the induced
oscillations is well defined by the mixing rate and is insensitive
to the initial conditions and the lattice size variations. The
observed behavior essentially differs from that predicted by the
Mean-Field model which is conservative.

\vspace{-0.09cm}
\par Complexity theory and associated methodologies are
transforming ecological research, providing new perspectives on old
questions as well as raising many new ones. Patterns and processes
resulting from interactions between individuals, populations,
species and communities in landscapes are the core topic of ecology.
Complex natural networks often share common structures such as
loops, trees and clusters, which contribute to widespread processes
including feedback, non-linear dynamics, criticality and
self-organization. Ecologists have long noted that the distribution,
abundance, and behavior of organisms are influenced by interactions
with other species. The complexity and the stability of ecosystems,
the study of food webs gained momentum in the late 1970s and early
1980s. Then new issues arise in ecology, such as environmental
change, spatial ecology and biodiversity. Experiments on pond food
webs show that the effects of species on ecosystem processes depend
on the interplay between environmental factors and trophic position.
In ecosystems, and in particular in food webs, the abovementioned
common structures have strong implications for their stability and
dynamics. Actually, a food web constitutes a special description of
a biological community with focus on trophic interactions between
consumers and resources~\cite{Rui05}. Food webs are deeply
interrelated with ecosystem processes and functioning since the
trophic interactions represent the transfer rates of energy and
matter within the ecosystem, in which the trophic webs are the
result of the interaction of different subgroups or \emph{modules}.
Recently an efficient \emph{dynamical clustering} (DC) algorithm for
identification of modules in complex networks, based on the
desynchronization properties of a given dynamical system associated
to the network was proposed in Ref.~\cite{Boc07}. Pluchino,
Rapisarda, and Latora in this issue (p. 395) apply the DC algorithm
to a well-known food web of marine organisms living in the
Chesapeake Bay, situated on the Atlantic coast of the United States.
By implementing the DC algorithm and using several dynamical
systems, such as R\"{o}ssler, Kuramoto, opinion changing rate model
(OCR), the authors are able to discover community configurations
with values of modularity higher than that calculated for the
reference configuration related to the main subdivision of the
Chesapeake Bay network in Benthic and Pelagic organisms. The DC
algorithm is able to perform, therefore, a very reliable
classification of the real communities existing in the Chesapeake
Bay food web.

\section{Bioinformatics}

Non protein-coding RNAs (ncRNAs) are a research hotspot in
bioinformatics, in fact recent discoveries have revealed new ncRNA
families performing a variety of roles, from gene expression
regulation to catalytic activities. In a very recent past, RNAs were
considered mere intermediates between the genome and the proteins.
Recent discoveries involving a variety of new ncRNA genes,
biological roles and action mechanisms have shown that the diversity
and importance of ncRNAs were underestimated. Non-coding RNA (ncRNA)
genes produce functional RNA molecules rather than encoding
proteins. Non-coding RNAs seem to be particularly abundant in roles
that require highly specific nucleic acid recognition without
complex catalysis, such as in directing post-transcriptional
regulation of gene expression or in guiding RNA
modifications~\cite{Edd01,Mac08}. Nowadays, it is known that
functional RNAs that do not code to proteins perform important
roles. They are involved in several cellular activities such as gene
silencing, replication, gene expression regulation, transcription,
chromosome stability, protein stability, translocation and
localization and RNA modification, processing and stability. RNA
regulatory secondary structures have been detected in almost all
living organisms ranging from viruses to Homo sapiens. The earliest
discoveries of their regulatory role have been performed in model
organisms such as the little worm C. elegans and in studies of the
interaction between plants and viruses. Small noncoding RNA
regulatory sequences are often characterized by the presence of
hairpin structures. A comprehensive study on $1832$ segments of
$1212$ complete genomes of viruses is presented here by Span\`{o},
Lillo, Miccich\`{e}, and Mantegna (p. 323). The authors show that in
viral genomes the hairpin structures of thermodynamically predicted
RNA secondary structures are more abundant than expected under a
simple random null hypothesis. The detected hairpin structures of
RNA secondary structures are present both in coding and in noncoding
regions for the four groups of viruses investigated in this
comprehensive study. For all groups hairpin structures of RNA
secondary structures are detected more frequently than expected for
a random null hypothesis in noncoding rather than in coding regions.
Differently from previous results obtained by the authors, they
observe that the detected hairpin structures are preferentially
located in the noncoding regions. However, at least in
herpesviruses, the degree of evolutionary conservation of these
structures is more pronounced in coding than in noncoding regions.

\vspace{-0.09cm}
\par The idea that bacteria are simple solitary creatures stems from
years of laboratory experiments in which they were grown under
artificial conditions. Under the demands of the wild, these
versatile life forms work in teams, in association and dynamic
communications. Bacteria can self-organize into hierarchically
structured colonies of $10^9$ to $10^{12}$ bacteria, each utilizing
a great variety of biochemical communication agents, such as simple
molecules, polymers, peptides, complex proteins, genetic material
and also viruses~\cite{Ben06,Ben05}. Ben-Jacob in this issue (p.
315) shortly reviews his far-reaching work on social behavior of
bacteria in colonies, guided by the assumption that they might shed
new light on the foundations and evolution of Biocomplexity. To face
changing environmental hazards, bacteria resort to a wide range of
cooperative strategies. They alter the spatial organization of the
colony in the presence of antibiotics for example. Bacteria form
complex patterns as needed to function efficiently. Bacteria modify
their colonial organization in ways that optimize bacterial
survival. Bacteria, Ben-Jacob argues, have collective memory by
which they track previous encounters with antibiotics. They
collectively glean information from the environment, communicate,
distribute tasks, perform distributed information processing and
learn from past experience. Bacteria are "smart" in their use of
cooperative behaviors that enable them to collectively sense the
environment. The author proposes a striking hypothesis: bacteria use
their genome computation capabilities and genomic plasticity to
collectively maintain exchange of meaning-bearing chemical messages
(semantic), and dialogues (pragmatic) for purposeful alteration of
colony structure and even decision-making, features that are
associated with intelligence. Collectively bacteria store
information, perform decision make decisions (e.g. to sporulate) and
even learn from past experience (e.g. exposure to antibiotics) –
features we begin to associate with bacterial social behavior and
even rudimentary intelligence. In other words, bacteria must be able
to sense the environment and perform internal information processing
for thriving on latent information embedded in the complexity of
their environment.

\par Symbolic sequences are investigated in many different
fields, including information theory, biological sequence analysis,
linguistics, chaotic time series, and communication theory. Many
efforts have been devoted to devise methods for generating
univariate or multivariate sequences with given statistical
properties~\cite{Tum07}. In the paper by Tumminello, Lillo, and
Mantegna (p. 333) in this issue, a method to generate multivariate
series of symbols from a finite alphabet with a given hierarchical
structure of similarities based on the Hamming distance is
introduced. The method presented here is based on a generating
mechanism that does not make use of mutation rate, which is widely
used in phylogenetic analysis. Interesting extensions of the
proposed method are the possibility of generating symbolic sequences
with correlations between different sites, useful to reproduce
dependencies between different sites of DNA, proteins, etc.., and of
assessing the role of the finite length of the series in discovering
the true phylogeny.

\section{Population Dynamics}

In the present-day context of global warming and habitat
destruction, there is an enhanced general interest in the impact of
environmental changes on biological populations evolution. A
simple-model population, whose individuals react with a certain
delay to temporal variations of their habitat, is presented in this
issue by Bena Coppex, Droz, Szwabi´nski, and P¸ekalski (p. 341). In
the case of a smooth variation of the environment, it was found
that, in general, for populations with small mutation amplitudes it
is more beneficial, in terms of the survival chance, to be
slow-reacting than to answer instantaneously to the variations of
the environment. However, for intermediate and large mutation
amplitudes, faster reactions are preferable to slower ones. In case
of a very-rapidly oscillating environment, the rapidity of reaction
influences only slightly the survival chances.

\vspace{-0.09cm}
\par The processes with some combination of randomness and periodicity
occur very often in nature. Life exists in the form of cycles, but
they never have exact period and amplitude. Chichigina (p. 347)
proposes in this issue a new model to describe the population cycles
in small rodents (lemming) of the north regions, like North America
and Siberian tundra. This model contains a noise source with memory.
Multiannual lemming density fluctuations are presented as a pulse
sequence, which correspond to the peaks of lemming density. The
memory is presented as some delay time after each pulse. Parameter
of periodicity, average period, correlation function and parameter
of synchronization are calculated for different places of North
America. Examples of equations modeling population dynamics of
lemmings (or their predators) are considered. A model of connected
oscillators gives the qualitative explanation of synchronization
effects and relation between synchronization and periodicity. This
model is also a very useful tool for modeling quasi-periodical
nature processes.

\vspace{-0.05cm}
\par The epidemic spread via a contact infection process in an
immobile population within the Susceptible-Infected-Removed (SIR)
model is analyzed in the paper by Naether, Postnikov, and Sokolov
(p. 353). This model describes a population consisting of three
kinds of individuals, namely the susceptible (S), the infected (I),
and the recovered/removed (R) ones. The transitions between these
states are governed by the infection transmission rate and the
characteristic recovery time. Both deterministic macroscopic
(through a partial differential equation (PDE)) and stochastic
microscopic (Monte-Carlo) approaches are applied. It is shown, that
the continuous description is not valid for small numbers of cell's
states. The results of Monte-Carlo simulations also reveal the
conditions of applicability of the PDE approach.

\vspace{-0.09cm}
\par The Verhulst model, which is a cornerstone
of empirical and theoretical ecology, is one of the classic examples
of self-organization in many natural and artificial
systems~\cite{Ciu96}. In the theoretical paper by Dubkov and
Spagnolo (p. 361), the transient dynamics of the Verhulst model
perturbed by arbitrary non-Gaussian white noise is investigated.
Based on the infinitely divisible distribution of the L\'{e}vy
process~\cite{Dub05}, the nonlinear relaxation of the population
density for three cases of white non-Gaussian noise is analyzed: (i)
shot noise; (ii) noise with a probability density of increments
expressed in terms of Gamma function; and (iii) Cauchy stable noise.
Exact results for the nonstationary probability distribution in all
cases investigated are obtained. For the Cauchy stable noise the
exact analytical expression of the nonlinear relaxation time is
derived. Due to the presence of a L\'{e}vy multiplicative noise, the
probability distribution of the population density exhibits a
transition from a trimodal to a bimodal distribution in asymptotics,
and the nonlinear relaxation time as a function of the Cauchy stable
noise intensity shows a nonmonotonic behavior with a maximum.

\par Population biologists use time-series data to infer the factors
that regulate natural populations and to determine when populations
may be at risk of extinction. Often, however, only a few of the
system's variables can be measured, while the rest of the variables
remain unobservable, or \emph{hidden}~\cite{Ion06}. Luchinsky,
Smelyanskiy, Millonas, and McClintock (p. 369) in their paper
addressed the problem of how to infer the unmeasured predator
dynamics, as well as the parameters of the nonlinear stochastic
dynamical models. As an example of how to solve a long-standing
ecological problem, the authors inferred an unobservable predator
trajectory, and parameter values, for a predator-prey model by
analysis of measurements of the prey dynamics that were (as is
typical) corrupted by noise.

\section{Motion of Proteins and Motor Protein}

An overview over recent studies on the model of Active Brownian
Motion (ABM)~\cite{Sch03} together with applications of ABM to
ratchets is given in the paper by Fiasconaro, Ebelin, and
Gudowska-Nowak (p. 403). The system has been studied under the
influence of smooth ratchet potentials: symmetrical (sinusoidal
potential) and not symmetrical (Mateos-type potential and tilted
ones). The system presents bifurcations of the asymptotic velocity
as a function of the energy transfer parameter, and three dynamical
regimes: 1) relaxation in a potential minimum (vanishing motion); 2)
oscillating motion in a well (limit cycle), and 3) flux motion with
two values of the asymptotic velocity. The motion in the flux regime
is then possible in two directions, even in the presence of a tilted
potential. The numerical simulations of the system under the action
of white Gaussian fluctuations show the effect of noise-controlled
directionality of the motion. Possible applications of the ABM
system in modelling molecular motors connected to the ATP synthesis/
hydrolysis have been briefly discussed~\cite{Bie07}.

\vspace{-0.03cm}
\par Over the past decades increasingly sophisticated techniques
have been developed to follow motor protein motion and manipulate
it. Processive motor proteins are tiny engines that utilize the
energy released in ATP hydrolysis to literally move in a
hand-over-hand fashion along a biopolymer~\cite{Bie07,Bie08}.
Kinesin, for instance, is a dimer, consisting of two units of about
350 amino acids, that moves along microtubule. Kinesin helps
maintain cell organization by pulling organelles and chemical-filled
vesicles to and from different parts of the cell. In the Bier's
paper (p. 415) a simple and intuitive model for chemo-mechanical
energy transduction is presented, that can quantitatively account
for back-stepping measured rates by experimentalists. The concept of
an overdamped Brownian stepper includes few adjustable parameters,
leads to a consistent accounting for the energy of ATP hydrolysis,
and makes some measured data derivable as implications of other
measured data.

\vspace{-0.03cm}
\par Protein molecules represent the final result of genetic expression,
and through their functions, they control key reactions in
ecological processes performed by microorganisms in aquatic,
terrestrial and certain artificial environments. Strategic analysis
of microbial proteins to elucidate microbial diversity and
ecosystem-level activities in the environment require an
appreciation of complexity inherent in protein
structure~\cite{Ogu05}. Based on observations that proteins can
readily be detected as components of dissolved organic matter, one
can use the protein analysis in ecology and environmental sciences
focusing on terrestrial ecosystems~\cite{Sch05}. Vibrational
spectroscopy is a powerful tool to study the functional activity of
proteins. In the experimental paper by Brandt, Chikishev,
Dolgovskii, Kargovskii, and Lebedenko (p. 419), the low-frequency
vibrational motions in proteins are analyzed, by discussing the
underlying physical mechanisms. The damping of these motions and the
effect of solvent molecules is experimentally studied by Raman
spectroscopy, by discussing their possible effect on the protein
functioning.

\section{Interdisciplinary Physics}

Although the functional role and precise physiological basis of
cardio-respiratory interactions are still being revealed, it has
been shown that this interaction changes between different states,
such as during an{\ae}sthesia. Phase transitions like phenomena in
synchronization have been shown to occur in rats as their depth of
anæsthesia varies. Transitions between different orders of
synchronization can potentially yield information about the
couplings and their evolution with time. Kenwright, Bahraminasab,
Stefanovska, and McClintock (p. 425) show in their paper in this
issue, that even during a steady state, such as in repose,
cardio-respiratory phase transitions exist and changes occur between
close orders of synchronization ratios. Exercise not only perturbs
the oscillators but also their interactions, which manifest as a
temporary reduction in synchronization. The cardiac and respiratory
frequencies were extracted from ECG and respiration signals by use
of marked events. The analysis was based on the phase dynamics
approach introduced by Kuramoto~\cite {Kur84} and its subsequent
applications to synchronization analysis. By using a phase-coupled
model with low-frequency noise, the authors show that
synchronization transitions are mainly due to the presence of
low-frequency fluctuations resulting from the activity of
lower-frequency oscillatory components.
\vspace{-0.1cm}

Although cancer is a leading cause of death in the world, it is
still little known about the mechanisms of its growth and
destruction. A detailed theoretical study on the mechanisms of
interaction between tumor tissue and immune system is necessary for
planning efficient strategies of treatment. In the paper by
Fiasconaro, Ochab-Marcinek, Spagnolo, and Gudowska-Nowak (p. 435) a
mathematical model describing the growth of tumor in the presence of
the immune response of a host organism is investigated. The model,
based on a reaction scheme representative of the catalytic
Michaelis-Menten scenario~\cite{Fia06}, is supplemented with
periodic treatment and external fluctuations in the tumor growth
rate. The resulting phenomenological equation modelling
cell-mediated immune surveillance against cancer exhibits
bistability. The two stationary points correspond to the state of a
stable tumor and the state of its extinction. A quantitative
analysis of mechanisms responsible for optimization of periodic
tumor therapy in the presence of spontaneous external noise is
given. Studying the behavior of the extinction time as a function of
the treatment frequency, the authors found the typical resonant
activation effect: For a certain frequency of the treatment, there
exists a minimum extinction time.

\vspace{-0.1cm} Physical and biological systems are continuously
perturbed by random fluctuations produced by noise sources always
present in open systems. Noise can be responsible for several
interesting and counterintuitive effects, such as resonant
activation (RA)~\cite{Doe92,Man00} and noise enhanced stability
(NES)~\cite{Agu01,Dub04}. Valenti, Augello, and Spagnolo (p. 443)
analyze the occurrence of these noise induced effects RA and NES in
the dynamics of a FitzHugh-Nagumo (FHN) system subjected to
autocorrelated noise, by finding meaningful modifications of these
phenomena due to the colored noise. In particular for strongly
correlated noise, the suppression of NES effect and persistence of
RA phenomenon, with an efficiency enhancement of the neuronal
response, is observed. The self-correlation of the colored noise
causes a reduction of the effective noise intensity, which appears
as a rescaling of the fluctuations affecting the FHN system.

\par The functionality of a complex biological system depends
on the correct exchange of information between the component parts.
In natural systems the environmental noise always affects the signal
that carries the information. Usually high levels of noise make
difficult to reveal signals, so that in everyday life the noise is
generally considered harmful in detecting and transferring
information. Under specific conditions, the noise can constructively
interacts with the system, so that effects induced by the noise,
such as stochastic resonance (SR), can improve the conditions for
signal detection. The experimental paper by Spezia, Curcio,
Fiasconaro, Pizzolato, Valenti, Spagnolo, Lo Bue, Peri, and Colazza
(p. 453) in this issue reports on experiments conducted on the
response of \emph{Nezara viridula} (L.) (Heteroptera Pentatonide)
individuals to sub-threshold signals. By investigating the role of
the noise in the vibrational communications occurring during the
mating behavior of \emph{N. viridula}, the authors find evidence of
stochastic resonance phenomenon~\cite{Fre02,Gam98}. In particular
the behavioural activation of the green bugs, described by the
Source-Direction Movement Ratio, has a nonmonotonic behaviour with a
maximum as a function of the noise intensity. This maximum
represents the optimal noise intensity which maximizes the
efficiency of the sexual communication between individuals of
\emph{N. viridula}. This behavior appears as the signature of the
soft threshold stochastic resonance~\cite{Gre04}. There is a
suitable noise intensity which maximizes the behavioral response of
the green bugs and this effect can be described by a soft threshold
model which shows the stochastic resonance
phenomenon~\cite{Dou93,Rus00}.

\par Almost all the processes that determine the properties of
the ocean take place at the sea surface and are related to the
interactions between the atmosphere and the ocean. The air-sea
fluxes of momentum, water and heat are of particular importance, in
relation to the patterns of currents, salinity and temperature. The
importance of the physical and chemical processes of the sea on the
reproductive biology of fishes (European anchovy) in the central
Mediterranean is shown in the experimental paper by Grammauta,
Molteni, Basilone, Guisande, Bonanno, Aronica, Giacalone, Fontana,
Zora, Patti, Cuttitta, Buscaino, Sorgente, and Mazzola (p. 459) in
this issue. The air-sea fluxes, the sea surface temperatures, and in
less extent the wind induced turbulence, may be considered proxies
of a favorable anchovy spawning sites, where adult specimens may
meet optimal environmental features for reducing the mortality of
the spawning products.

\section{Conclusions}

The present issue aimed to gather together scientists studying
different ecological structures, with different points of view and
backgrounds, ranging in space and time scales from microscopic
bacterial growth, to cancer growth and to fish dynamics and
exhibiting temporal regular and irregular oscillations which may be
due to the intrinsic dynamics or caused by external sources and
modulated by noise. We hope that the readers of this issue should
assess the current status of this research field, should be able to
determine (a) the problems to be attacked theoretically and
numerically, and (b) the best theoretical and numerical models to be
used in each particular case. We expect that the experimental
physicists receive new suggestions and ideas for their research by
reading this special issue. From experimental and theoretical points
of view recent advances in the fields of bacterial growth, RNA
structure in virus genome, tumor development, social behavior in
animal groups, competing species in various spatial geometries,
epidemics spreading, and other ecological topics have been presented
in this topical issue. Specifically the following specific topics
were addressed: self-organization in ecosystems, social behavior of
bacteria, cooperative phenomena, noise induced transitions,
patchiness, clustering and heterogeneity, effects of initial and
boundary conditions, external control of pattern formation, noise
induced phenomena and stochastic dynamics of out of equilibrium
systems. A broad range of new results on ecological complex systems
are reported, establishing the state of the art of the field. The
workshop "Ecological Complex Systems: Stochastic Dynamics and
Patterns" has succeeded in providing a forum for the exchanges of
ideas between experimentalists, theoreticians and modellers and
cross-fertilization between the different ecological topics, by
contributing essentially in this way to the development of the
field.

\subsection*{Acknowledgements}
We wish to acknowledge the European Science Foundation by STOCHDYN
programme, the University of Palermo, the "Dipartimento di Fisica e
Tecnologie Relative" of the Palermo University, and the "Azienda
Autonoma Provinciale per l'incremento turistico (AAPT)" of the
Regional Province of Palermo's city, for supporting the 2007
Workshop on \emph{Ecological Complex Systems}.

Moreover, we wish to warmly thank all the authors and referees who
have contributed, in different way, to the preparation of this
topical issue. Our warmest gratitude to the Editor in Chief of the
\emph{European Physical Journal B}, Prof. Frank Schweitzer for his
continuous assistance and advice in the management of the issue, and
to Mr. Nicolas Puyaubreau for his constant and unfailing support
during the long editorial procedure.

%\newpage


\begin{thebibliography}{}

\bibitem{Sci99}
See the special section on "\emph{Complex Systems}", Science
\textbf{284}, 79 (1999)

\bibitem{Sci01}
See the special section on "\emph{Ecology Through Time}", Science
\textbf{293}, 623 (2001)

\bibitem{Fre02}
JA Freud, L. Schimansky-Geier, B. Beisner, A. Neiman, DF Russell, T.
Yakusheva and F. Moss, \emph{Behavioral StochastcResonance: How the
noise from a Daphnia swarm enhances individual prey capture by
juvenile paddlefish}, J. Theor. Biol. \textbf{214}, 71 (2002)

\bibitem{Spa04}
B. Spagnolo, D. Valenti and and A. Fiasconaro, \emph{Noise in
Ecosystems: A Short Review}, Math. Biosc. and Eng. \textbf{1}, 185
(2004)

\bibitem{Ebe05}
Werner Ebeling and Bernardo Spagnolo, \emph{Noise in Condensed
Matter and Complex Systems}, Fluctuation and Noise Letters
\textbf{5} (2), L159-L161 (2005)

\bibitem{Chi05}
O. Chichigina, D. Valenti, and B. Spagnolo, \emph{A Simple Noise
Model with Memory for Biological Systems}, Fluctuation and Noise
Letters, \textbf{5} (2), L243-L250 (2005)

\bibitem{Spa08}
Bernardo Spagnolo and Alexander Dubkov, \emph{Critical Phenomena and
Diffusion in Complex Systems}, Int. J. of Bifurcation and Chaos
\textbf{18} (9), 2643 (2008)

\bibitem{Liu08}
Quang-Xing Liu, Zhen Jin and Bai-Lian Li, \emph{Resonance and
frequency-locking phenomena in spatially extended
phytoplankton-zooplanbkton system with additive noise and periodic
forces}, J. Stat. Mech. \textbf{5}, P0511 (2008)

\bibitem{Lev04}
H. Levine and E. Ben Jacob, \emph{Physical schemata underlying
biological pattern formation: examples, issues and strategies},
Physical Biology \textbf{1}, 14 (2004)

\bibitem{Ben06}
E. Ben Jacob, Y. Shapira and AI. Tauber, \emph{Seeking the
foundation of cognition in bacteria: From Schroedinger's negative
entropy to latent information}, Physica A \textbf{359}, 495 (2006)

\bibitem{Kha06}
E. Khain and L.M. Sander, \emph{Dynamics and pattern formation in
invasive tumor growth}, Phys. Rev. Lett. \textbf{96}, 188103 (2006)

\bibitem{The02}
G. Theraulaz, E. Bonabeau, SC. Nicolis, RV. Sole, V. Fourcasie, S.
Blanco, R. Fournier, JL. Jolly, P. Fernandez, A. Grimal, P. Dalle
and JL. Deneubourg, \emph{Spatial patterns in ant colonies}, Proc.
Nat. Acad. Sci. USA \textbf{99}, 9645 (2002)

\bibitem{Cou05}
ID Couzin, J. Krause, NR. Franks and SA Levin, \emph{Effective
Leadership and decision making in animal groups on the move}, NATURE
\textbf{433}, 513 (2005)

\bibitem{Erd05}
U. Erdmann, W. Ebeling and AS. Mikhailov, \emph{Noise induced
transition from translational to rotational motion of swarms}, Phys.
Rev. E \textbf{71}, 051904 (2005)

\bibitem{Pek06}
A.Pekalski and M. Droz, \emph{Self-organised packs selection in
predator-prey ecosystems}, Phys. Rev. E \textbf{73}, 021913 (2006)

\bibitem{San03}
L.M. Sander, C.P. Warren and I.M. Sokolov, \emph{Epidemics, disorder
and percolation}, Physica A \textbf{325}, 1 (2003)

\bibitem{Gro06}
T. Gross, CJD d' Lima and B. Blasius,\emph{ Epidemic dynamics on an
adaptive network}, Phys. Rev. Lett. \textbf{96}, 208701 (2006)

\bibitem{Kha05}
E. Khain, L.M. Sander and AM. Stein, \emph{A model of glioma
growth}, Complexity \textbf{11}, 53 (2005)

\bibitem{Fia06}
Alessandro Fiasconaro and Bernardo Spagnolo, Anna Ochab-Marcinek and
Ewa Gudowska-Nowak, \emph{Co-occurrence of resonant activation and
noise-enhanced stability in a model of cancer growth in the presence
of immune response}, Phys. Rev. E \textbf{74}, 041904(10) (2006)

\bibitem{Sok05}
I.M. Sokolov, R. Metzler, K. Pant, and M.C. Williams, \emph{Target
Search of N Sliding Proteins on a DNA}, Biophys. J. \textbf{89},
895-902 (2005)

\bibitem{Wan04}
J. Wang, \emph{The Complex Kinetics of Protein Folding in Wide
Temperature Ranges}, Biophys. J. \textbf{87}, 2164-2171 (2004)

\bibitem{Li08}
Da-wei Li, Haijun Yang, Li Han  and Shuanghong Huo, \emph{Predicting
the Folding Pathway of Engrailed Homeodomain with a Probabilistic
Roadmap Enhanced Reaction-Path Algorithm}, Biophys. J. \textbf{94},
1622-1629 (2008)

\bibitem{Aba03}
E. Abad, A. Provata and G. Nicolis, \emph{Reactive dynamics on
fractal sets: Anomalous fluctuations and memory effects}, Europhys.
Lett. \textbf{ 61}, 586 (2003).

\bibitem{Cam01}
S. Camazine, J.L. Deneubourg, N.R. Franks, J. Sneyd, E. Bonabeau, G.
Theraulaz, \emph{Self-organization In Biological Systems} (Princeton
University press, 2001)

\bibitem{Zha99}
V.P. Zhdanov, \emph{Surface restructuring and kinetic oscillations
in heterogeneous catalytic reactions}, Phys. Rev. E \textbf{60},
7554 (1999)

\bibitem{Rui05}
P.C. de Ruiter, V. Wolters, J.C. Moore, K.O. Winemiller, \emph{Food
web ecology: playing Jenga and beyond}, Science \textbf{309}, 68
(2005)

\bibitem{Boc07}
S. Boccaletti, M. Ivanchenko, V. Latora, A. Pluchino, A. Rapisarda,
D\emph{etecting complex network modularity by dynamical clustering},
Phys. Rev. E \textbf{75}, 045102(R) (2007)

\bibitem{Edd01}
S.R. Eddy, \emph{Non–coding RNA genes and the modern RNA world},
Nature Reviews Genetics \textbf{2}, 919 (2001)

\bibitem{Mac08}
A. Machado-Lima, H. A. del Portillo, A. M. Durham,
\emph{Computational methods in noncoding RNA research}, J. Math.
Biol. \textbf{56}, 15 (2008).

\bibitem{Ben05}
E. Ben-Jacob, Y. Aharonov, Y. Shapira, \emph{Bacteria harnessing
complexity}, J. Biofilm \textbf{1}, 239 (2005)

\bibitem{Tum07}
M. Tumminello, F. Lillo, R.N. Mantegna, \emph{Hierarchically nested
factor model from multivariate data}, Europhys. Lett. \textbf{78},
30006 (2007)

\bibitem{Ciu96}
S. Ciuchi, F. de Pasquale, B. Spagnolo, \emph{Self Regulation
Mechanism of an Ecosystem in a Non Gaussian Fluctuation Regime},
Phys. Rev. E \textbf{54}, 706 (1996)

\bibitem{Dub05}
Alexander A. Dubkov and Bernardo Spagnolo, \emph{Generalized Wiener
Process and Kolmogorov's Equation for Diffusion Induced by
Non-Gaussian Noise Source}, Fluctuation and Noise Letters
\textbf{5}(2), L267-L274 (2005)

\bibitem{Ion06}
E.L. Ionides, C. Breto, A.A. King, \emph{Inference for nonlinear
dynamical systems}, Proc. Nat. Acad. Sci. 103, 18438 (2006)

\bibitem{Sch03}
Frank Schweitzer, \emph{Brownian Agents and Active Particles,
Springer Series in Synergetics}, Springer, Berlin 2003.

\bibitem{Bie07}
M. Bier, \emph{The stepping motor protein as a feedback control
ratchet}, Biosystems \textbf{88}, 301 (2007)

\bibitem{Ogu05}
O.A. Ogunseitan, \emph{Microbial Diversity} (Blackwell Publishing,
Oxford, 2005)

\bibitem{Sch05}
W.X. Schulze, \emph{Protein analysis in dissolved organic matter:
What proteins from organic debris, soil leachate and surface water
can tell us - a perspective}, Biogeosciences \textbf{2}, 75 (2005)

\bibitem{Bie08}
Martin Bier, \emph{The energetics, chemistry, and mechanics of a
processive motor protein}, Biosystems \textbf{93}, 23-28 (2008)

\bibitem{Kur84}
Y. Kuramoto, \emph{Chemical Oscillations, Waves, and Turbulence}
(Springer-Verlag, Berlin, 1984)

\bibitem{Doe92}
C.R. Doering, J.C. Gadoua, \emph{Resonant activation over a
fluctuating barrier}, Phys. Rev. Lett. 69, 2318 (1992)

\bibitem{Man00}
R.N. Mantegna, B. Spagnolo, \emph{Experimental Investigation of
Resonant Activation}, Phys. Rev. Lett. 84, 3025 (2000)

\bibitem{Agu01}
N. Agudov and B. Spagnolo, \emph{Noise enhanced stability of
periodically driven metastable states}, Phys. Rev. Rap. Comm. E
\textbf{64}, 035102(R) (2001)

\bibitem{Dub04}
A. A. Dubkov, N. V. Agudov and B. Spagnolo, \emph{Noise enhanced
stability in fluctuating metastable states}, Phys. Rev. E
\textbf{69}, 061103(7) (2004)

\bibitem{Gam98}
L. Gammaitoni, P. H\"{a}nggi, P. Jung, F. Marchesoni,
\emph{Stochastic resonance}, Rev. Mod. Phys. \textbf{70}, 223 (1998)

\bibitem{Gre04}
P.E. Greenwood, U.U. M\"{u}ller, L.M. Ward, \emph{Soft threshold
stochastic resonance}, Phys. Rev. E \textbf{70}, 051110 (2004)

\bibitem{Dou93}
J.K. Douglass, L. Wilkens, E. Pantazelou, and F. Moss, \emph{Noise
enhancement of information transfer in crayfish mechanoreceptors by
stochastic resonance}, Nature 365, 337-340 (1993).

\bibitem{Rus00}
D. F. Russel, L. A. Wilkens And F. Moss, \emph{Use of Behavioural
Stochastic Resonance by Paddle Fish for feeding}, Natures 402,
291-294 (2000).

\end{thebibliography}
\end{document}